\documentclass[prl,twocolumn,nofootinbib, showpacs,preprintnumbers,amsmath,amssymb]{revtex4-1}
\usepackage{dcolumn}
\usepackage{bm}
\usepackage[hidelinks=true]{hyperref}
\usepackage{comment}
\usepackage{amsmath}
\usepackage{amssymb}
\usepackage{amsfonts}
\usepackage[mathscr]{eucal}
\usepackage{color}
\usepackage{xcolor}
\usepackage[active]{srcltx}
\usepackage[mathscr]{eucal}
\usepackage{enumitem}

\newcommand{\be}{\begin{equation}}
\newcommand{\ee}{\end{equation}}
\newcommand{\bea}{\begin{eqnarray}}
\newcommand{\eea}{\end{eqnarray}}
\newcommand{\bse}{\begin{subequations}}
\newcommand{\ese}{\end{subequations}}
\newcommand{\beqa}{\begin{eqnarray}}
\newcommand{\eeqa}{\end{eqnarray}}
\newcommand{\beqar}{\begin{eqnarray*}}
\newcommand{\eeqar}{\end{eqnarray*}}
\newcommand{\bi}{\begin{itemize}}
\newcommand{\ei}{\end{itemize}}
\newcommand{\bn}{\begin{enumerate}}
\newcommand{\en}{\end{enumerate}}

\newcommand{\ba}{\begin{array}}
\newcommand{\ea}{\end{array}}
\newcommand{\bc}{\begin{center}}
\newcommand{\ec}{\end{center}}

\definecolor{darkred}{rgb}{0.9,0.05,0.05}
\definecolor{darkblue}{rgb}{0.05,0.05,0.6}
\definecolor{darkgreen}{rgb}{0.05,0.6,0.05}
\definecolor{brightgreen}{rgb}{0.1,0.9,0.1}

\renewcommand*{\eqref}[1]{%
	\begingroup
	\hypersetup{
		linkcolor=darkblue,
		linkbordercolor=darkblue,
	}%
	\textcolor{darkblue}{(\ref{#1})}%
	\endgroup 
}

\hypersetup{linkcolor=red,citecolor=brightgreen,urlcolor=black,colorlinks=true}

\newcommand{\mrd}{\mathrm{d}}
\newtheorem{definition}{Definition}
\newtheorem{proposition}{Proposition}
\begin{document}
	
	\title{
	 Redundant and Physical Black Hole Parameters:\\
	{\normalsize
	{Is there an independent physical dilaton charge?}} }
	
	\author{K. Hajian and M.M. Sheikh-Jabbari}

	\affiliation{\vspace*{3mm} \textit{School of Physics, Institute for Research in Fundamental
			Sciences (IPM), P.O.Box 19395-5531, Tehran, Iran} \vspace*{2mm}}
	\vfil
	
	\begin{abstract}
	\noindent
	Black holes as solutions to gravity theories, are generically identified by a set of parameters. Some of these parameters are associated with black hole physical conserved charges, like ADM charges. There can also be some ``redundant parameters.'' We propose necessary conditions for a parameter to be physical. The conditions are essentially integrability and non-triviality of the charge variations arising from ``parametric variations,"  variation of the solution with respect to the chosen parameters. In addition, we prove that variation of the redundant parameters which do not meet our criteria do not appear in the first law of thermodynamics. As an interesting application, we show that dilaton moduli are redundant parameters for black hole solutions to Einstein-Maxwell-(Axion)-Dilaton theories, because variations in dilaton moduli would render entropy, mass, electric charges or angular momenta non-integrable. Our results are in contrast with modification of the first law due to scalar charges suggested in Gibbons-Kallosh-Kol paper \cite{Gibbons:1996af} and its follow-ups. We also briefly discuss implications of our results for the attractor behavior of extremal black holes.  
	\end{abstract}

	\maketitle


\section{Introduction}

Solutions to generic physical theories are specified by a set of parameters. These parameters are generically integrals of motion that are specified by initial and boundary conditions. 
Although there could be cases where some of these parameters take discrete-values, here we will only focus on real-valued parameters which can be varied continuously and define the {``solution space''} as the space of solutions spanned by these  parameters. 

All physical observables associated with a solution are functions of these parameters. In particular, among the physical observables there are conserved charges. The celebrated Noether theorem \cite{Noether}, while relating the conserved charges to symmetries of the theory, provides the functional form of conserved charges on this solution space.

In diffeomorphism invariant gravity theories, where we do not necessarily have (globally defined) time-like Killing vectors, the notion of ``conservation'' should be handled with special care. In addition, application of usual Noether theorem may face various challenges (e.g. see \cite{Compere-thesis} and references therein for a review). To tackle these issues various different proposals and formulations have been proposed. However, there are aspects of this problem which still remain as a matter of debate to date.

As it may generically happen, physical observables may only be functions of a subset of parameters spanning the solution space. Or in other words, a part of the solution parameters may not appear in any physical observable. This can, in particular, happen in theories with local gauge symmetry (like diffeomorphisms in gravity) or in theories with field redefinition symmetry at the level of classical action. For example,  some of the solution parameters could be gauge artifacts which may be removed in  different coordinate systems or by a choice of gauge. It may also happen that the theory enjoys a field redefinition symmetry and some of the parameters may be  related to a choice of dynamical fields to describe the system.  An important example, which we consider and analyze here, is the shift symmetry in systems with a dialton field. One would hence face the question which of the solution parameters are really physical ones.

An answer to this question, which is implicitly used in the literature, can be the following: any parameter that can be removed by a symmetry transformation (transformations which do not change the equations of motion and a given boundary condition) is redundant, while the parameters which appear explicitly in the conserved charges like mass, angular momentum etc. are physical. This inaccurate resolution has shortcomings in the following two situations. Firstly, it is now an established fact that there are some diffeomorphisms or gauge transformations to which one can  associate non-trivial conserved charges \cite{Brown:1986nw}. In such cases, we are dealing with family of diffeomorphic, but still distinct, geometries \cite{CHSS, Compere:2015knw,Sheikh-Jabbari:2016lzm,Hawking:2016sgy, Afshar} (or in generic case, gauge equivalent, but still distinct solutions, e.g. see \cite{QED-symmetries}) which are specified by a number of arbitrary functions. These families of solutions may hence be labeled by infinite parameters which are e.g. related to the Fourier modes of  these functions and there is a well-defined charge associated with each of these parameters. Secondly, there can be parameters removable by the action of a symmetry, while they appear explicitly in the well-established conserved charges like mass etc. An important example of such a parameter is the dilaton modulus in the dilaton shift symmetry. 

While the method and algorithm we provide can be used in a wider context, here we would tackle the question described above for a specific class of solutions to gravity theories, the black holes. We mainly focus  on the family of stationary black holes, those which have a time-like Killing vector outside their (event) horizon. It is now established that black holes generically Hawking radiate, a black body radiation emitted from  any thermal system at the Hawking temperature and there should be an entropy associated with them \cite{Hawking:1976rt,Bekenstein:1973ft}. It is  also established that  black holes obey laws of thermodynamics \cite{Bardeen:1973gd,Iyer:1994ys}. Black hole thermodynamical quantities, which are either the extensive conserved charges, or the intensive (chemical) potentials, are all functions over the {black hole solution space}. Our goal here is to provide  unambiguous criteria and algorithm to distinguish the physical and redundant parameters of these solutions.

\section{Physical vs. redundant solution parameters}
Let us start by crystallizing the definition of solution parameters described in the introduction.  
\vspace*{-0.1cm}    
\begin{definition}\label{def solution par}
	$\!\!\!$\textbf{.} Given a Lagrangian density $\mathcal{L}$ and a solution to its e.o.m's with a given boundary condition, in some specific gauge and coordinate system, ``solution parameters" $p_i$ are constants in dynamical fields each of which can be varied while e.o.m's are still satisfied.  
\end{definition}
This definition makes a clear distinction between solution parameters and conserved charges attributed to a solution; solution parameters and conserved charges are conceptually different entities, which may or may not be related. To keep the distinction in mind, we denote the set of parameters by $p_i$, while the standard notation mass $M$, angular momentum $J$, electric charge $Q$, entropy $S$ etc. is used for the conserved charges. Notice that this definition covers parameters which may appear in a solution and correspond to the ``residual symmetries and charges''  \cite{Sheikh-Jabbari:2016lzm} discussed in the introduction. On the other hand, it excludes parameters of the theory, constants which appear explicitly in the Lagrangian, like the Newton constant $G$ or the cosmological constant $\Lambda$. Moreover, the definition clarifies that these parameters cannot be constrained or related to each other through {equations of motion (e.o.m)} or boundary conditions.

\subsection{Charges vs. charge variations and the integrability}
Solution parameters can be related to the conserved charges of a solution through symmetries of the theory and/or solution. Conserved charges may be calculated by different methods.  One can recognize two classes of such methods: those which provide a prescription to calculate the charges directly, and the methods which calculate charge \emph{variations} first, and if integrable, then the finite charges. 
Noether charge \cite{Noether,Banados:2016zim}, Komar charge \cite{Komar:1958wp}, ADM method \cite{ADM} and its later developments like Brown-York \cite{Brown:1992br}, and ADT methods \cite{ADT}, are examples in the first category (reviewed e.g. in \cite{Szabados}). In these methods, the conserved charges are read directly from the solution (or possibly its subtraction off a reference solution). As examples of  methods in the second category, {quasi-local method \cite{Kim:2013zha}}, covariant phase space formulation  \cite{Compere-thesis, Ashtekar:1987hia,Crnkovic:1987at,Lee:1990gr,Wald:1993nt,Iyer:1994ys,Wald:1999wa,Barnich:2001jy, BC} and solution phase space method (SPSM) \cite{HS:2015xlp} can be mentioned. 

The methods based on charge variations, especially in the context of gravity, are more precise, applicable to a wider range of solutions and may be uniformly applied to solutions with various  asymptotic behavior. In these methods integrability of charge variations may yield non-trivial constraints on physical observables. The main idea we propose in this paper is to use the integrability conditions to distinguish between physical and redundant solution parameters. We employ SPSM which we find the more rigorous method among those in the charge variation class. Here, we briefly highlight the main ingredients and features of this method. For more details, the reader can refer to \cite{HS:2015xlp} or  \cite{Hajian:2016kxx,Ghodrati:2016vvf}. 

\subsection{Solution Phase Space Method, a quick review}

SPSM elaborates on the connection between solution parameters and conserved charges. To calculate a charge variation four inputs are needed: 1) A  Lagrangian $\mathcal{L}$ on $d$ dimensional spacetime with coordinates $x^\mu$; {2) A symmetry to which the charge variation is attributed; 3) A solution to the e.o.m of the theory specified with dynamical field configuration $\Phi$ (e.g. metric $g_{\mu\nu}$, gauge field $A_\mu$, scalar field $\phi$, etc.);} 4) A perturbation around the solution $\delta \Phi(x^\mu)$ satisfying linearized e.o.m. 

{Based on covariant phase space formulation \cite{Ashtekar:1987hia,Lee:1990gr,Iyer:1994ys,Barnich:2001jy, BC},}  SPSM combines the four inputs in a simple relation, to introduce a charge variation $\delta H_\epsilon$,
 
{\small{
		\begin{equation}\label{cov charge def}
		\delta H_\epsilon\equiv \int_{\Sigma} \boldsymbol{\omega}(\delta\Phi,\delta_\epsilon\Phi;\Phi)= \oint_{\partial\Sigma} \boldsymbol{k}_\epsilon(\delta\Phi;\Phi)\,,
		\end{equation}}
}
\vspace*{-0.3cm}

\noindent where $\Sigma$ is a (codimension one) Cauchy surface and $\partial \Sigma$ is its (codimension two) boundary. $\delta H_\epsilon$ is conserved if it is independent of the choice of $\Sigma$. 
The $d-1$-form $\boldsymbol{\omega}$ (called symplectic current) is  on-shell {closed} ($d\boldsymbol{\omega}=0$ on-shell) and its form is determined through the Lagrangian  $\mathcal{L}$, {e.g. see  \cite{Iyer:1994ys,Barnich:2001jy, BC}.} To write the second equation we have used $\boldsymbol{\omega}=d \boldsymbol{k}$ on-shell and the Stokes' theorem. Explicit form of $\boldsymbol{k}$ for generic Lagrangians  may be found e.g. in \cite{Ghodrati:2016vvf}. 

Information of the symmetry is in $\epsilon$, {which is in general} a combination of a diffeomorphism generator vector field $\xi^\mu$ and a gauge transformation $\lambda$, denoted by $\epsilon=\{\xi^\mu,\lambda\}$ \cite{Barnich:2003xg,HS:2015xlp}. They act on fields as $\delta_\epsilon\Phi\equiv\mathscr{L}_\xi\Phi+\delta_\lambda A$ where $\mathscr{L}_\xi$ is Lie derivation and $\delta_\lambda A_\mu=\partial_\mu \lambda$.  

{In SPSM, $\epsilon$ is taken to} be some specific subset of general diffeomorphisms and gauge transformations, called \emph{symplectic symmetries}, for which

\vspace*{-0.2cm}
{\small
	\begin{equation}\label{symp symm 1}
	\boldsymbol{\omega}(\delta\Phi,\delta_\epsilon\Phi,\Phi)= 0\,,
	\vspace*{-0.5cm}	\end{equation}}

\noindent over the specified set of solutions $\Phi$ and $\delta \Phi$. This nice feature makes the conservation to be guaranteed and renders the conserved charge variations to be independent of the codimension-2 integration surface $\partial\Sigma$. Therefore, the charges can be obtained from integrating $\boldsymbol{k}_\epsilon$ over any smooth and closed codimension-2 surface $\mathscr{S}$ inside the bulk which encompass any non-smoothness, singularity or closed-time-like-curves of the solution \cite{HS:2015xlp}.

{Among all  solutions of the theory, SPSM focuses on those which may be denoted by $\Phi(x^\mu; p_i)$ where $p_i$ are the parameters discussed in definition \ref{def solution par}.
For the last but not least input, the standard condition which is imposed on $\delta\Phi$ is that it satisfies linearized equations of motion.}\footnote{As has been emphasized in the literature (e.g. see \cite{Szabados,Banados:2016zim}) and references therein, to guarantee the conservation and finiteness of the charge variations, one needs to impose carefully prescribed boundary conditions on the field variations $\delta\Phi$. However, one may show that the symplectic symmetry condition \eqref{symp symm 1} relaxes the strict choices of the falloff conditions on $\delta\Phi$ discussed in the literature; for symplectic symmetries  $\delta\Phi$ may asymptotically be as large as the background solution $\Phi$. {Consequently, one need not be concerned with the falloff behavior of parametric variations associated with symplectic symmetries.}} {Here} we consider a more specific set of field variations, the \emph{parametric variations} \cite{HSS:2014twa, HS:2015xlp}. 
Given a set of solutions  $\Phi(x^\mu;p_i)$ parametric field variations are defined as  

\vspace*{-0.2cm}
{\small
	\begin{equation}\label{para variation}
	\hat \delta \Phi=\frac{\partial{\Phi}}{\partial p_i}\delta p_i\,.
	\end{equation}}

\vspace*{-0.2cm}
\noindent It is clear from the above definition that parametric variations satisfy linearized equations of motion.  To emphasize that the variations we will be considering hereafter are just parametric variations we will denote them by $\hat \delta X$. In particular  parametric charge variations will be denoted by $\hat \delta H_\epsilon$, which is obtained by surface integrals over {$\boldsymbol{k}_\epsilon(\hat\delta\Phi;\Phi)$}. \footnote{We would like to emphasize that the examples of symmetries and charges associated with non-trivial gauge transformations \cite{CHSS,Compere:2014cna,Afshar,Compere:2015knw} also fall into the parameteric charge variations \eqref{para variation}.}

\noindent \textbf{Integrability.} As mentioned, for symplectic symmetries and after integration over the closed codimension 2 surface $\mathscr{S}$, we obtain the conserved charge variation  $\hat\delta H_\epsilon$ which is independent of $\mathscr{S}$ and may be viewed as a one-form {over the solution space:} $\hat\delta H_\epsilon=F^i\delta p_i$, where $F^i$'s are some functions of $p_i$'s, not the coordinates $x^\mu$. One may then ask if there exists a function $H_\epsilon (p_i)$ variation of which gives $\hat\delta H_\epsilon$. In other words,
 
 	\vspace*{-0.2cm}
 	{\small
 		\begin{equation}\label{del H F(p)}
 		\hat\delta H_\epsilon=F^i\delta p_i\stackrel{?}{=}\hat\delta (H_\epsilon (p_i))\,.
 		\vspace*{-0.3cm} 	\end{equation}}
 
\vspace*{-0.2cm}
\noindent \textbf{Exact symmetries and their charges.}  An important set of symplectic symmetries which we will focus on in this work are \emph{exact symmetries} \cite{Barnich:2003xg,HS:2015xlp}. Denoting them by $\eta=\{\zeta^\mu,\lambda\}$, they are defined through {$\delta_\eta\Phi=0$ over the class of $\Phi(x;p_i)$  solutions}. Therefore, $\zeta^\mu$ is necessarily a Killing vector. The nice property of exact symmetries is that the corresponding charge variations are free of ambiguities which generically appear in the definition of $\boldsymbol{k}_\eta$ \cite{HS:2015xlp}.
Hereafter we will focus on  $\hat\delta H_\eta$, parametric charge variations associated with exact symmetry $\eta$ and analyze integrability of the associated charge variation.

In SPSM putting the diffeomorphisms and gauge transformations together in the generators is not just a convention. It is mandatory because of integrability of charges. For example, consider  a stationary and axi-symmetric black hole solution, with the coordinate adopted such that the Killing vectors to be $\partial_t$ and $\partial_\varphi$ respectively. Generically, exact symmetries for mass, angular momentum, electric charge and entropy of the horizon $\mathrm{H}$, respectively denoted by $M$, $J$, $Q$ and $S_{_\mathrm{H}}$, are \cite{Iyer:1994ys,HS:2015xlp,Hajian:2016kxx,Ghodrati:2016vvf}
	
	\vspace*{-0.5cm}
	{\small
		\begin{align}\label{generators}
		&\eta_{_M}=\{\partial_t+\varOmega_{_\infty}\partial_\varphi, -\varPhi_{_\infty}\}, \quad \eta_{_J}=\{-\partial_\varphi,0\},\nonumber\\
		&\eta_{_{S_\mathrm{H}}}=\frac{2\pi}{\kappa_{_\mathrm{H}}}\{\partial_t+\varOmega_{_\mathrm{H}}\partial_\varphi,-\varPhi_{_\mathrm{H}}\}, \quad \eta_{_Q}=\{0,1\},
		\end{align}}
	
	\vspace*{-0.4cm} 
	\noindent where $\varOmega$, $\varPhi$ and $\kappa$ denote angular velocity, electric potential and surface gravity on the horizon or at spatial infinity and $\eta_{_{S_\mathrm{H}}}$ is the vector field generating the Killing horizon. Noting that these coefficients  are functions of parameters $p_i$, their specific presence in corresponding $\eta$'s (in particular  $1/\kappa_{_\mathrm{H}}$ factor in front of $\eta_{_{S_\mathrm{H}}}$) is crucial for the integrability of charges.   Note also that in the context of SPSM  entropy and other exact symmetry charges are symplectic and hence may be calculated over arbitrary $\mathscr{S}$ in the bulk. 
	
As a final remark we note that	in the SPSM setup, the first law of black hole thermodynamics is simply deduced from quasi-local version of a \emph{local} identity which relates $\eta_{_{S_{\mathrm{H}}}}$ to the generators of other charges. For instance, for the generators in \eqref{generators} this identity is explicitly
	
	\vspace*{-0.5cm}
	{\small
		\begin{equation}
		\frac{\kappa_{_\mathrm{H}}}{2\pi}\eta_{_{S_{\mathrm{H}}}}=\eta_{_M}-(\varOmega_{_\mathrm{H}}-\varOmega_{_\infty})\eta_{_J}-(\varPhi_{_\mathrm{H}}-\varPhi_{_\infty})\eta_{_Q}\,.
		\end{equation}}
	
	\vspace*{-0.5cm}
	\noindent Then, by $T_{_\mathrm{H}}\equiv\frac{\kappa_{_\mathrm{H}}}{2\pi}$ and the linearity of $\delta H_\epsilon$ \eqref{cov charge def} in $\epsilon$, 
	
	\vspace*{-0.5cm}
	{\small
		\begin{equation}\label{first law}
	{T_{_\mathrm{H}}}\delta S_{_\mathrm{H}}=\delta M-(\varOmega_{_\mathrm{H}}-\varOmega_{_\infty})\delta J-(\varPhi_{_\mathrm{H}}-\varPhi_{_\infty})\delta Q\,,
		\end{equation}}
	\vspace*{-0.5cm}
	
	\noindent for any field variations $\delta\Phi$ which satisfy linearized equation of motion, including $\hat \delta \Phi$s.

\subsection{Necessary conditions for a black hole solution parameter to be physical }

Solution parameters can be divided into two families, physical and redundant parameters. For the class of black hole solutions, we propose the following two necessary conditions for a solution parameter $p$ to be physical:
\textit{
\begin{enumerate}
\item all black hole conserved charge parametric variations appearing in the first law should be integrable with respect to $p$,\label{condition 1}
\item there should exist at least one non-zero integrable conserved charge variation with respect to $p$.\label{condition 2}
\end{enumerate}}
 Physical motivations for the first condition is coming from standard thermodynamics and that charge variations in the first law should be integrable. As for the second condition, for a parameter to be physical it is expected to have non-trivial contribution to some of conserved charges which may or may not appear in the first law. Based on the two conditions, we prove the following proposition.
\begin{proposition}
$\!\!\!$\textbf{.}	Variations of redundant solution parameters which are rejected by either of the two conditions do not appear in the first law of thermodynamics.
\end{proposition}
\noindent\textit{Proof.} By the assumption,  a {rejected} solution parameter  $p$ fails at least one of the conditions \textit{\ref{condition 1}}. or \textit{\ref{condition 2}}. If it fails the condition \textit{\ref{condition 1}}., $\delta p$ should be put to zero, in order to recover the integrability of charges in the first law. Then, trivially the first law can not be extended to include variations proportional to $\delta p$. If $p$ fails the condition \textit{\ref{condition 2}}, there does not exist any integrable charge variation including $\delta p$ to be added to the first law.  \hfill $\Box$ 
	
Notice that each one of the conditions \textit{\ref{condition 1}.} and \textit{\ref{condition 2}.} has a harmless ambiguity. In the condition \textit{\ref{condition 1}.} one can always be skeptical about proper choice of the generators. In other words, the loss of integrability might be because of taking an incorrect generator. In the condition \textit{\ref{condition 2}.} a parameter might not contribute to the known charges, while itself can introduce a new conserved charge. We assume that both of these ambiguities can be overcome by further investigation of any given  black hole solution.

\section{Dilaton moduli are redundant solution parameters} 
In gravity theories obtained from (Kaluza-Klein) compactification and reduction of a higher dimensional gravity theories we generically have ``moduli fields''  which are generically related to geometry of the compactification manifold \cite{moduli}. Moduli are generically scalar fields whose background or asymptotic value is not fixed by the field equations.  Presence of moduli in the solutions denotes the freedom in the compactification manifold which are not dynamically fixed. This feature called moduli problem precludes obtaining physical theories with falsifiable prediction in lower dimensions \cite{moduli-fixing}.    

A specific example of such moduli is coming from dilaton field(s) in theories of (super)gravity. Dilaton is a scalar field $\phi$ with the property that the Lagrangian remains invariant under the $\phi\to \phi+\phi_0,$ for a constant $\phi_0$  (possibly together with rescaling of other fields). Therefore, for any given solution to these theories, we have a freedom in the asymptotic value of the dilaton field, the dilaton modulus $\phi_\infty$. This free constant fits well in Definition \ref{def solution par}, so it is a solution parameter. Interestingly or intriguingly, if we compute the conserved charges, e.g. using ADM method or Komar integrals, they show explicit dependence on the value of $\phi_\infty$. As a result, it has been considered to be a physical parameter, e.g. see \cite{Gibbons:1996af, Gibbons-1982} and papers citing them. In particular, it has been argued that $\delta \phi_\infty$ should also appear in the first law. Given our criteria and discussions of the last section, we show that the dilaton modulus is a redundant (unphysical) parameter. We demonstrate this for two typical black hole solutions to two typical, but very simple, dilatonic gravity theories. 
\vskip2mm
\noindent\textbf{1$^{st}$ example: Einstein-Dilaton gravity.} Consider  4-dimensional Einstein-dilaton gravity 

\vspace*{-0.3cm}
{\small
	\be\label{Ein-Dil-Lag}
	\mathcal{L}=\frac{e^{-\phi} }{16 \pi G}(R-2\nabla_\mu\phi\nabla ^\mu\phi).
	\ee}

\vspace*{-0.3cm}
\noindent The Kerr geometry \cite{Kerr:1963ud} is a black hole solution to this theory with the metric and dilaton fields

\vspace*{-0.5cm}
{\small
	\begin{align}\label{Kerr metric}
	&\mathrm{d}s^2= \Big(-(1\!-\! f)\mathrm{d}t^2+\frac{\rho ^2}{\Delta}\mathrm{d}r^2+{\rho ^2} \mathrm{d}\theta ^2 -2 fa\sin ^2 \theta\,\mathrm{d}t \mathrm{d}\varphi\nonumber\\
	&\hspace*{0.5cm}+\left(r^2+a^2+fa^2\sin ^2\theta \right)\sin ^2\theta\,\mathrm{d}\varphi ^2\Big), \qquad\quad   \phi=\phi_\infty,\\
	&\rho^2 \equiv r^2+a^2 \cos^2 \theta\,,\quad \Delta \equiv r^2+a^2-2mr\,,\quad
	f\equiv\frac{2mr}{\rho ^2}\,,\nonumber
	\end{align}
}

\vspace*{-0.5cm}
\noindent which has three solution parameters $p_i\in\{m,a,\phi_\infty\}$. For the black hole solution \eqref{Kerr metric}

{\small 
	\vspace*{-0.4cm}\begin{equation}\label{Kerr Omega kappa}
	\varOmega_{_\mathrm{H}}\!=\!  \frac{a}{2m r_{_\mathrm{H}} }, \qquad  \varOmega_{_\infty}\!=\!0, \qquad   \kappa_{_\mathrm{H}}=\frac{\sqrt{m^2-a^2}}{2m r_{_\mathrm{H}} }\,,\ \  
	\end{equation}}

\vspace*{-0.3cm}
\noindent where $r_{_\mathrm{H}}=m+\sqrt{m^2-a^2}$.

Let us now check  which  of the parameters pass our physicality  criteria. As for  the condition \textit{\ref{condition 1}}., we need to calculate parametric variations of mass,  angular momentum, and (at least) one of the entropies of the horizons. By the SPSM and for the generators in \eqref{generators} without any gauge field contribution (see \cite{Ghodrati:2016vvf} to find $\boldsymbol{k}_\epsilon$ for the chosen Lagrangian), the final results are

\vspace*{-0.3cm}
{\small
	\begin{align}\label{ex 1 var}
	&\hspace*{-0.3cm} \hat \delta M=\frac{e^{-\phi_\infty}}{G}\Big(\delta m- \frac{m}{2}\delta 
	\phi_\infty\Big)\,,\\
	&\hspace*{-0.3cm} \hat \delta J=\frac{e^{-\phi_\infty}}{G}\Big(a\delta m+m \delta a-ma\delta \phi_\infty\Big),\\
	&\hspace*{-0.3cm} \hat \delta S_{_\mathrm{H}}=\frac{e^{-\phi_\infty}}{G}\Big( \frac{\partial(\frac{A_{\mathrm{H}}}{4})}{\partial m}\delta m+\frac{\partial(\frac{A_{\mathrm{H}}}{4})}{\partial a}\delta a-\frac{A_{\mathrm{H}}}{4}\delta \phi_\infty\Big)
	\end{align}}

\vspace*{-0.4cm}
\noindent where $A_{\mathrm{H}}=4\pi(r_{_\mathrm{H}}^2+a^2)=8\pi m r_{_\mathrm{H}}$. It is clear that in the charges above, mass is not integrable unless $\delta\phi_\infty=0$. By the failure of condition \textit{\ref{condition 1}}. $\phi_\infty$ is a redundant solution parameter, and should not be varied. Setting $\delta\phi_\infty=0$ all charges become integrable, yielding
\footnote{We note that changing the normalization of the generator of mass $\eta_{_\mathrm{M}}\to e^{\frac{\phi_\infty}{2}}\eta_{_\mathrm{M}}$ makes the variation of mass in \eqref{ex 1 var} integrable. Then, finite mass would be $\tilde {M}=\frac{m}{G}e^{-\frac{\phi_\infty}{2}}$. This is an example of the harmless ambiguity in the condition \textit{\ref{condition 1}}. This possibility is not acceptable because it leads a first law not in the standard form: $T_{_\mathrm{H}}\delta S=e^{-\frac{\phi_\infty}{2}}\delta \tilde{M}-\varOmega_{_\mathrm{H}}\delta J$.}

\vspace*{-0.3cm}
{\small
	\begin{equation}\label{Ein-dilaton BH charges}
	M= \frac{me^{-\phi_\infty}}{G}, \qquad J=\frac{mae^{-\phi_\infty}}{G},\qquad  S_{_\mathrm{H}}=\frac{A_\mathrm{H}e^{-\phi_\infty}}{4G}.
	\end{equation}}

\vspace*{-0.3cm}
\noindent One may readily verify that the above together with \eqref{Kerr Omega kappa} satisfy the first law in \eqref{first law} without any contribution from $\delta \phi_\infty$. By non-vanishing contribution of $\delta m$ and $\delta a$ to the charges, $m$ and $a$ pass the criterion \textit{\ref{condition 2}.} too. Therefore, $m$ and $a$ might be considered as physical parameters.

\vskip2mm
\noindent\textbf{2$^{nd}$ example: Einstein-Maxwell-Dilaton theory.}
As the  second example, we choose a typical dilaton black hole which has been first introduced in \cite{Garfinkle:1990qj,Preskill:1991tb} (see also \cite{Kallosh:1992ii,Goulart:2016cuv}). The Lagrangian of the four dimensional theory with metric $g_{\mu\nu}$, a scalar field $\phi$, and the Maxwell gauge field $F=\mrd A$ as dynamical fields is

\vspace*{-0.1cm}
{\small
\begin{equation}\label{EMD-Lag}
\mathcal{L}=\frac{1}{16\pi G}\Big(R-e^{-2\phi}F_{\mu\nu}F^{\mu\nu}-2\nabla_\mu \phi \nabla^\mu \phi\Big)\,.
\end{equation}}

\vspace*{-0.3cm}
\noindent For simplicity and to be specific we consider the spherically symmetric black hole solution 

\vspace*{-0.3cm}
{\small
\begin{align}\label{dilaton sol}
&\mrd s^2=-f\mrd t^2+\frac{\mrd r^2}{f}+(r^2-\Sigma^2)(\mrd \theta^2+\sin^2 \theta \mrd \varphi^2)\,, \nonumber\\
& f=\frac{(r-r_+)(r-r_-)}{r^2-\Sigma^2},\quad \Sigma=\frac{-q^2}{2m}, \quad r_\pm=m\pm r_0 \nonumber\\
& A=\frac{qe^{\phi_\infty}}{r-\Sigma}\mrd t,\quad \phi=\frac{1}{2}\log \left(\frac{r+\Sigma}{r-\Sigma} \right)+\phi_\infty,
\end{align}}

\vspace*{-0.3cm}
\noindent where $r_0=m+\Sigma$. This black hole has three solution parameters $p_i\in\{m,q,\phi_\infty\}$ and its (outer) horizon is located at $r=r_+$. For this solution in  the chosen coordinates and gauge, $\varOmega_{_\mathrm{H}}\!=\!\varOmega_{_\infty}\!=\!\varPhi_{_\infty}\!=0$, and

\vspace*{-0.4cm}
{\small
\begin{align}\label{Chemicals-ED-Kerr}
\varPhi_{_\mathrm{H}}=\frac{q\,e^{\phi_{\infty}}}{2m},\qquad \kappa_{_\mathrm{H}}=\frac{1}{4m}\,.
\end{align}}

\vspace*{-0.3cm}
\noindent By the SPSM and the generators \eqref{generators} (see \cite{Ghodrati:2016vvf} for $\boldsymbol{k}_\epsilon$ for Lagrangian \eqref{EMD-Lag}), parametric variations of the three conserved charges are \footnote{{Interested readers can find explicit calculations in a supplementary Mathematica code available \href{https://drive.google.com/file/d/0B8AqjYFfBKw0ZUc3OFVDT3lVdUE/view?usp=sharing}{\underline{in this link}}.} }

\vspace*{-0.4cm}
{\small
\be\begin{split}
\hat \delta M=\frac{1}{G}\delta m-&\frac{q^2}{2Gm}\delta \phi_\infty,  \quad\! \hat \delta Q=\frac{e^{-\phi_\infty}}{G}(\delta q-q\,\delta \phi_\infty),\\
&\hat \delta S_{_\mathrm{H}}=\frac{8\pi m }{G}\delta m-\frac{4\pi q}{G}\delta q.
\end{split}\ee	
}

\vspace*{-0.3cm}
\noindent The results clearly show that the {mass is integrable only when variation of modulus parameter  $\delta \phi_\infty$ is set to zero.} By the failure of condition \textit{\ref{condition 1}.} the solution parameter $\phi_\infty$  is not physical. Integrating the charge variations we obtain

\vspace*{-0.5cm}
{\small
\begin{equation}\label{dilaton BH charges}
 M= \frac{m}{G}, \qquad Q=\frac{qe^{-\phi_\infty}}{G},\qquad  S_{_\mathrm{H}}=\frac{\pi(4m^2-2q^2)}{G},
\end{equation}}

\vspace*{-0.5cm}
\noindent {with appropriate choices of reference points \cite{HS:2015xlp}.} They satisfy the first law in \eqref{first law} with $\delta \phi_\infty=0$. Our result is in contrast with the inclusion of variation of $\delta\phi_\infty$ in the first law,  as suggested in \cite{Gibbons:1996af} or it follow ups. 

\section{Discussion}

So far, we have demonstrated through two examples that dilaton moduli would lead the charges of a black hole non-integrable and hence the dilaton modulus is a redundant parameter. Here we discuss more about the physical significance and implications of this result.
\vskip2mm
\noindent{\textbf{How generic is the redundancy of dilaton moduli?}}\vskip2mm
\noindent {Although we have shown redundancy of dialton moduli for very specific and simple theories, we argue below that our results are expected to hold for more general cases. Considering the first example, Einstein-dilaton theory has the following field redefinition shift symmetry}
 
 \vspace*{-0.2cm}
 {\small
 	\begin{equation}\label{shift-sym-metric}
 	\phi\to \phi+\phi_0. 
 	\end{equation}}
 
 \vspace*{-0.3cm} 
 \noindent {The dilaton modulus $\phi_\infty$ is a solution parameter associated with this symmetry. However, value of $\phi_\infty$, as explicitly seen from the Lagrangian \eqref{Ein-Dil-Lag} and also from \eqref{Ein-dilaton BH charges}, can be reabsorbed into the definition of the Newton constant (see comment V. below). So, $\phi_\infty$ is expected to be a redundant parameter.}

Similarly in our second example, the Lagrangian \eqref{EMD-Lag} is invariant under the transformations

\vspace*{-0.2cm}
{\small
	\begin{equation}\label{shift-sym-gauge-field}
	\phi\to \phi+\phi_0\,, \quad\quad A_\mu \to e^{\phi_0}A_\mu\,. 
	\end{equation}}

\vspace*{-0.3cm}
\noindent The dilaton modulus $\phi_\infty$ is a solution parameter that appears in the solution \eqref{dilaton sol} by this symmetry transformation. Hence, any choice of $\phi_\infty$  corresponds to a choice of the scale for  the electric charge, absorbed into the vacuum electric permittivity parameter, usually set to one. However, the choice of scale for $A_\mu$ should not change physical contents of the system. So, $\phi_\infty$ is expected to be redundant. 

We finish this part by the following remarks:
\vspace*{-0.2cm}\begin{itemize}[leftmargin=0.35cm]
\item[I.] Introducing other solution parameters, e.g. one for magnetic monopole, would not change the outcome of our realization of redundancy of $\phi_\infty$, because failure of integrability on a subset of parameters still yields  failure of condition \textit{\ref{condition 1}}.
\item[II.] Integrability is a property which is insensitive to any \emph{fiducial reparametrization} of solution parameters $p_i\to f_i (p_j)$ in which $f_i$'s are some functions with non-zero Jacobian determinant. So, our criterion rejects $\phi_\infty$ to be physical, irrespective of the chosen parametrization.
\item[III.] As a side result of redundancy of  $\phi_\infty$, $\phi_\infty$ or the so-called ``scalar charges''  \cite{Gibbons:1996af,Rasheed:1997ns,scalar-charge} do not appear in the Smarr relation. This is in agreement with statements in  \cite{Gibbons:1996af,Rasheed:1997ns,scalar-charge}.
\item[IV.] Although here we discussed  theories with one dilaton modulus field, it is very easy to see that the same analysis, arguments and hence results, hold for generic cases with several dilaton or axion moduli.
\item[V.] Our conclusion for not varying $\phi_\infty$ (unlike other physical parameters) is in accord with the discussions in \cite{Goldstein, Mann-et-al}, where it is argued that $\phi_\infty$ (and the value of moduli in general) should be viewed as parameters of the theory, rather than solution. For example  choice of $\phi_\infty$ may be absorbed into the definition of Newton constant $G$ in the Einstein-dilaton theory \eqref{Ein-Dil-Lag}, and into the definition of vacuum permittivity (which is conventionally set to one)  in the Einstein-Maxwell-dilaton theory \eqref{EMD-Lag}. Therefore, we should not vary the moduli when first law of black hole thermodynamics is considered. 
\item[VI.] Besides black holes there are other solutions to gravity theories for which one can define charges, charge variations and laws of (thermo)dynamics. Near Horizon Extremal Geometries (NHEG) are among the most extensively studied of such geometries, e.g. see \cite{HSS:2013lna, HSS:2014twa,CHSS,HS:2015xlp}. One can readily verify that our ``parameter redundancy criteria'' and the analysis here extends to the family of NHEG.
\end{itemize}

\noindent{\textbf{Implications for attractor mechanism.}} It has been observed that value of dilaton at the horizon $\phi_{{_\mathrm{H}}}$ for extremal black hole solutions, and hence the entropy, are independent of $\phi_\infty$ and that $\phi_{{_\mathrm{H}}}$ is completely determined by the value of physical charges \cite{Ferrara:1995ih,Ferrara:1996dd,Strominger:1996kf}. This phenomenon is called attractor mechanism. In view of our discussions on the redundancy of $\phi_\infty$, we would like to revisit the attractor behavior for extremal black holes. It is readily seen (and is a well known fact) that appearance of $\phi_\infty$ as a solution parameter  is an artifact of the dilaton shift symmetry.

Although our discussions below hold for more general cases with several moduli fields and more complicated actions,  for simplicity  consider a case with one dilaton modulus and denote black hole with parameters $p_i\in\{\bar p,\phi_\infty\}$ for $i-1$ number of physical parameters $\bar p$. Using the dilaton shift symmetry, e.g. those given in \eqref{shift-sym-metric} and \eqref{shift-sym-gauge-field}, it is always possible to write the solution such that\footnote{To see this, first put $\phi_\infty=0$ everywhere in an arbitrary given parametrization, and then reintroduce $\phi_\infty$ back to the solution via the symmetry transformations.}

\vspace*{-0.4cm}
{\small
\begin{align}\label{dilaton nice rep}
\hspace*{-3mm}\phi(x^\mu;p_i)\!=\!\bar\phi(x^\mu;\bar p)+\phi_\infty,\quad \Phi(x^\mu;p_i)\!=\!f_\Phi({\phi_\infty})\bar \Phi(x^\mu;\bar p),
\end{align}}

\vspace*{-0.5cm}
\noindent for an appropriate function $f_\Phi(\phi_\infty)$, where in the above $\Phi$ denotes any other field than the moduli.  This parametrization is convenient because
\vspace*{-0.2cm}\begin{itemize}[leftmargin=0.35cm]
\item[(1)] the metric, possibly up to an overall scale factor, is independent of moduli and therefore, horizon radius $r_{_\mathrm{H}}$ is only a function of $\bar p$. Moreover, the horizon area and hence the entropy (after a possible scaling in the Newton constant) is independent of the moduli. Note that  these features are true for both extremal or {non-extremal} black hole solutions. 
\item[(2)] Extremality condition, which is read from the metric, is a constraint  only among the $\bar p$'s, and can be denoted as $C(\bar p)=0$. 
\item[(3)] As a result of the above, 

\vspace*{-0.4cm}
{\small
	\begin{align}\label{phi ext rH}
	\phi_{_\mathrm{H}}=\bar \phi_{_\mathrm{H}}+\phi_\infty.
	\end{align}}

\vspace*{-0.5cm}
\noindent
where $\bar \phi_{_\mathrm{H}}$ is a function of just $\bar p$, not the $\phi_\infty$. 
\end{itemize}

Let us now revisit the attractor mechanism. Although one can always choose a parametrization, using specific fiducial parameters $p_i$, such that $\phi_\infty$ do not explicitly appear in the conserved charges like mass, angular momentum and electric charge, e.g. in \eqref{dilaton BH charges} $q\to qe^{-\phi_\infty}$, this leads to moduli dependence in the entropy. This is a parametrization often used in the papers discussing the attractor behavior in the literature. In this parametrization $\phi_\infty$ also enters the extremality condition $C(\bar p)\to \tilde C(\bar p,\phi_\infty)$, providing a constraint between $\phi_\infty$ and $\bar p$'s. 
One may use this latter extremality constraint to undo the superficial moduli dependence of the entropy.

In a similar manner, $\phi_{_\mathrm{H}}$ for extremal black holes may have $\phi_\infty$ dependence  in an arbitrary parametrization (different than the one used in \eqref{dilaton nice rep} and \eqref{phi ext rH}). Nonetheless, our arguments above clarify that one can always write $\phi_{_\mathrm{H}}$ as $\bar\phi_{_\mathrm{H}}$ which is a function of $\bar p$,  plus $\phi_\infty$. So, in parametrization \eqref{dilaton nice rep} $\phi_{_{\mathrm{H}}}$ explicitly depends on $\phi_{\infty }$ for \emph{extremal} and non-extremal black holes.

\textbf{Acknowledgements:} We would like to thank members of quantum gravity group in IPM, especially Mohammad  Vahidinia and Prieslei Goulart for helpful discussions. This work has been supported by the \emph{Allameh Tabatabaii} Prize Grant of \emph{National Elites Foundation} of Iran and the \emph{Saramadan grant} of the Iranian vice presidency in science and technology and the junior research chair in black holes of the Iranian NSF. MMShJ also thanks ICTP Net-68 scientific network project.

\section{Appendix: Review of covariant phase space formulation}

Here we briefly review covariant phase space formulation \cite{Ashtekar:1987hia,Crnkovic:1987at,Lee:1990gr,Wald:1993nt,Iyer:1994ys,Wald:1999wa,Barnich:2001jy, Compere-thesis, Ghodrati:2016vvf,Hajian-Seraj-thesis}).  The formulation begins with a given covariant Lagrangian top form $\mathbf{L}\equiv\star\mathcal{L}$ with dynamical fields $\Phi(x^\mu)$ in $d$-dimensional spacetime which generically include metric, gauge fields and other matter fields. 
The generator for a combination of a diffeomorphism and a gauge transformation will be denoted by $\epsilon\equiv\{\xi^\mu,\lambda\}$, where $\xi^\mu$ and $\lambda$ are some vector and scalar field respectively. Infinitesimal transformation generated by the $\epsilon$ is $\delta_\epsilon\Phi=\mathscr{L}_\xi\Phi+\delta_\lambda \Phi$ in which $\mathscr{L}_\xi$ denotes the Lie derivation, and $\delta_\lambda \Phi_\mu$ the gauge transformation. Charge variation associated with $\epsilon$, denoted by $\delta H_\epsilon$, is calculated by the following integration over a Cauchy surface $\Sigma$

\vspace*{-0.2cm}
{\small
	\begin{equation}\label{charge vol}
	\delta H_\epsilon\approx\int_{\Sigma} \delta \mathbf{\Theta}(\delta_\epsilon\Phi,\Phi)-\delta_\epsilon \mathbf{\Theta}(\delta\Phi,\Phi)\,,
	\end{equation}
}

\vspace*{-0.3cm}
\noindent such that the $\delta$ in the first term does not act on the $\epsilon$. The sign $\approx$ is used to denote the equality \emph{on-shell}. The $\Phi$ is a solution to field equations and $\delta\Phi$ denotes perturbation around it which satisfy linearized e.o.m. The $\mathbf{\Theta}$, which is a $d-1$-form, is the surface term appearing in the variation of Lagrangian

\vspace*{-0.2cm}
{\small
	\begin{equation}\label{def Theta}
	\delta \mathbf{L}=\mathbf{E}_\Phi\delta\Phi+\mrd \mathbf{\Theta}(\delta\Phi,\Phi)\,,
	\end{equation}}

\vspace*{-0.4cm} 
\noindent where all of the e.o.m's are denoted by $\mathbf{E}_\Phi$.

 The integrand in \eqref{charge vol} is called symplectic current, denoted by $\boldsymbol{\omega}(\delta\Phi,\delta_\epsilon\Phi,\Phi)$, and is bilinear in $\delta \Phi$ and $\delta_\epsilon\Phi$. This $d-1$-form is  closed on-shell $\mrd\boldsymbol{\omega}\approx 0$, which entails two features: 1) the current $\boldsymbol{\omega}$ is an exact form on-shell $\boldsymbol{\omega}(\delta\Phi,\delta_\epsilon\Phi,\Phi)\approx \mrd \boldsymbol{k}_\epsilon(\delta\Phi,\Phi)$, 2) it satisfies continuity equation. By the first feature and Stokes' theorem, charge variations can be calculated as a surface integral over the boundary $\partial \Sigma$

\vspace*{-0.2cm}
{\small
	\begin{equation}\label{charge surf}
	\delta H_\epsilon\approx\oint_{\partial\Sigma} \boldsymbol{k}_\epsilon(\delta\Phi,\Phi)\,.
	\end{equation}}

\vspace*{-0.2cm}
\noindent Notice that equation above is linear in $\epsilon$. To see how to find the $\boldsymbol{k}_\epsilon$ and its explicit form for general enough theories, one can refer to e.g. \cite{Ghodrati:2016vvf}. The second feature yields conservation, i.e. independence of $\delta H_\epsilon$ from the choice of $\Sigma$, if there is not any flux of current across the boundary $\partial\Sigma$. To achieve this latter, and generally to identify the set of allowed $\Phi$ and $\delta\Phi$, some fall-off conditions at spatial or null asymptotics are imposed. This set of allowed solutions $\Phi$ constitute a phase space manifold $\mathcal{M}$ with the $\delta\Phi$'s in its tangent space. The symplectic form of this phase space is $\boldsymbol{\omega}(\delta_1\Phi,\delta_2\Phi,\Phi)=\delta_1\mathbf{\Theta}(\delta_2\Phi,\Phi)-\delta_2\mathbf{\Theta}(\delta_1\Phi,\Phi)$ where $\delta_1\Phi,\delta_2\Phi$ are two arbitrary perturbations. 

To find finite conserved charges one needs to integrate $\delta H_\epsilon$ in \eqref{charge surf} over the $\mathcal{M}$, on a curve connecting a reference field $\Phi_0$ to a chosen field $\Phi$. Not all of the $\epsilon$'s have integrable or conserved charges. The integrability condition, i.e. independence of $H_\epsilon$ from the choice of the mentioned curve, turns out to be \cite{Wald:1999wa,Compere:2015knw}

\vspace*{-0.4cm}
{\small
	\begin{equation}\label{integrability cond modified}
	\hspace*{-0.2cm}\oint\displaylimits_{\partial\Sigma} \Big(\xi\cdot \boldsymbol{\omega}(\delta_1\Phi,\delta_2\Phi,\Phi)+\boldsymbol{k}_{\delta_1\epsilon}(\delta_2\Phi,\Phi) -\boldsymbol{k}_{\delta_2\epsilon}(\delta_1\Phi,\Phi)\Big)\approx 0\,,
	\end{equation}}

\vspace*{-0.3cm}
\noindent for all choices of $\Phi$ and $\delta_{_{1,2}}\Phi$ in $\mathcal{M}$ and its tangent bundle. 

 There are three types of ambiguities in the formulation. The first one is ambiguity in definition of Lagrangian $\mathbf{L}\to\mathbf{L}+\mrd \boldsymbol{\mu}$  for an arbitrary $d-1$-form $\boldsymbol{\mu}$. It can be shown that this ambiguity does not enter in the definition of charges in \eqref{charge vol} \cite{Iyer:1994ys}. Second ambiguity is $\mathbf{\Theta}\to \mathbf{\Theta}+\mrd \mathbf{Y}$ in the definition of $\mathbf{\Theta}$ through \eqref{def Theta}. The last ambiguity is in $\boldsymbol{k}_\epsilon\to \boldsymbol{k}_\epsilon+\mrd \mathbf{Z}_\epsilon$ in the definition of $\boldsymbol{k}_\epsilon$. $\mathbf{Y}$-term generally affects the conserved charges through $\boldsymbol{k}_\epsilon\to \boldsymbol{k}_\epsilon+\delta \mathbf{Y}(\delta_\epsilon\Phi)-\delta_\epsilon \mathbf{Y}(\delta \Phi)$, while the $\mathbf{Z}_\epsilon$ cannot, because by Stokes' theorem it is integrated over boundaries of $\partial \Sigma$. There is an interesting and important set of symmetries, the symplectic symmetries for which $\boldsymbol{\omega}(\delta\Phi,\delta_\epsilon\Phi,\Phi)\approx 0$ \cite{CHSS,HS:2015xlp}. For exact symmetries which are a subset of symplectic symmetries, $\mathbf{Y}$ ambiguity does not contribute to  $\boldsymbol{k}_\epsilon$ because its contributions is linear in the vanishing variations $\delta _\epsilon\Phi=0$ \cite{HS:2015xlp}.

\end{document}